\documentclass[
 reprint,              
 amssymb,       
 aps,                   
 prd,                   
]{revtex4-2}

\usepackage[T1]{fontenc}
\usepackage[dvipsnames]{xcolor}
\usepackage{dcolumn}
\usepackage{bm}
\usepackage{comment}
\usepackage{amsmath}
\usepackage{mathtools}
\usepackage{graphicx}
\usepackage{silence}
\usepackage{aas_macros}
\usepackage{placeins}
\usepackage{slantsc} 
\usepackage{booktabs}
\usepackage[hidelinks,colorlinks=true,linkcolor=blue,citecolor=green]{hyperref}
\usepackage{ulem}

\definecolor{brightpink}{rgb}{1.0, 0.1, 0.4}

\def\vmax{\ensuremath{V_{\rm max}}}
\def\rmax{\ensuremath{R_{\rm max}}}
\def\cvir{\ensuremath{c_{200}}}
\def\mvir{\ensuremath{M_{200}}}
\def\rvir{\ensuremath{R_{200}}}

\def\simlt{\stackrel{<}{{}_\sim}}

\newcommand{\code}[1]{\texttt{\detokenize{#1}}}

\newcommand{\cactus}{\code{CaCTus}}
\newcommand{\nexus}{{\sc nexus+}}
\newcommand{\Nbody}{{\textit{N}--body}}

\newcommand{\unit}[1]{\ensuremath{\mathrm{\,#1}}\xspace}
\newcommand{\unitlogicspace}[2]{%
  \ifthenelse{\isempty{#1}}%
    {\unit{#2}}
    {\ensuremath{{{#1}\, \unit{#2}}}}
  }

\newcommand{\Msun}[1][]{#1\,h^{-1}\,M_\odot}
\newcommand{\Mpch}[1][]{#1\,h^{-1}\,\mathrm{Mpc}}

\defcitealias{Hellwing2021}{H21}
\defcitealias{2025A&A...700A..65M}{Paper~I}
\defcitealias{2014MNRAS.442.2271S}{SCP14}
\defcitealias{2017MNRAS.466.4974M}{M17}
\DeclareUnicodeCharacter{2212}{\ensuremath{-}}

\begin{document}

\preprint{APS/123-QED}

\title{Cosmic-web environmental impacts on the halo substructure boosts to dark matter annihilation signals}
\author{Feven Markos Hunde}%
 \email{fevenm@cft.edu.pl}
\author{Wojciech A. Hellwing}
\email{hellwing@cft.edu.pl}
\author{Maciej Bilicki}

\affiliation{Center for Theoretical Physics, Polish Academy of Sciences, Aleja Lotników 32/46, 02-668 Warsaw, Poland}

\date{\today}
\begin{abstract}
The annihilation of dark matter (DM) particles is expected to produce Standard Model particles, providing a potential indirect signature of DM. The clumpy substructure of DM haloes amplifies the expected annihilation signal, an effect commonly quantified by the subhalo boost factor. Standard semi-analytic models usually treat this boost as a universal function of host-halo mass, neglecting systematic variations induced by the large-scale environment. In this work, we extend this framework by incorporating the influence of the cosmic web on subhalo populations. Using simulation-calibrated, environment-dependent ratios for host-halo concentrations, the subhalo mass function, and internal-structure proxies of subhalos based on the $V_{\max}$--$R_{\max}$ relation, we compute environment-conditioned boost predictions for haloes residing in filaments, walls, and voids. Our main result is the boost factor at fixed host-halo mass, expressed relative to the cosmic-mean prediction, $B(M,\mathrm{env})/B_{\mathrm{CM}}(M)$. We find a clear environmental modulation: in the fiducial distance-dependent model, filament haloes show a mass-dependent transition from a $\sim 15\%$ suppression at the low-mass end to a modest enhancement of $\sim 12\%$ for massive hosts, wall haloes remain intermediate, while void haloes stay suppressed by roughly $30$--$33\%$ across the explored host-mass range. These results should be interpreted as deterministic model predictions obtained by propagating environment-dependent ingredient ratios through two standard semi-analytic boost frameworks. We provide an environment-aware prescription for subhalo boosts, together with modular environmental corrections that may also be useful in indirect-detection forecasts, strong-lensing mass modeling, and related halo-population applications.
\end{abstract}

\maketitle
\section{\label{sec:intro} Introduction}
Cosmological and astrophysical observations indicate that dark matter (DM) constitutes over $80\%$ of the total matter in the Universe \citep{2018RvMP...90d5002B, 2020A&A...641A...6P}. DM plays a crucial role in shaping the formation and evolution of cosmic structures, yet its fundamental particle nature remains unknown. Numerous theoretical candidates have been proposed, ranging from macroscopic objects such as primordial black holes \citep{2021arXiv211002821C} to particle candidates including Weakly Interacting Massive Particles (WIMPs) \citep{1996PhR...267..195J, 2005PhR...405..279B, 2012PDU.....1..194B, Roszkowski_2018}, axions \citep{2006ARNPS..56..293A, 2016PhR...643....1M}, sterile neutrinos \citep{2010ARA&A..48..495F, 2021arXiv210900767K}, and others.

Experimental searches span a wide mass range, from ultralight axions in the sub-eV regime to WIMPs at the GeV--TeV scale, but have so far yielded no conclusive detections \citep{2016JPhG...43a3001M, 2017arXiv170704591B}. This lack of direct detection has motivated complementary approaches that seek to identify potential astrophysical signatures of DM interactions.

One such promising avenue is the indirect detection of DM. This approach searches for gamma rays, neutrinos, or other Standard Model particles produced by DM annihilation or decay in astrophysical environments. The expected flux depends on the underlying DM distribution, typically scaling with the square of the density \citep{2014MNRAS.442.2271S, 2017MNRAS.466.4974M, 2019Galax...7...68A}. Consequently, dense and clumpy regions, such as the galactic center, dwarf spheroidal galaxies, and galaxy clusters, are considered particularly promising targets for current and next-generation experiments \citep{2006A&A...455...21C, 2010PhRvD..81d3532K, 2012PDU.....1..194B, 2025arXiv250922609P}. 

Within the cold dark matter (CDM) paradigm, cosmic structures grow hierarchically through gravitational instability, in a bottom-up scenario where the first objects to collapse correspond to the highest initial density fluctuations \citep{Bond1996}. These small haloes merge and accrete to form progressively larger systems \citep[][]{White1978, 2005MNRAS.364.1105S}. This process naturally produces a population of subhalos within larger haloes. The most massive subhalos host luminous dwarf satellites, while many lower-mass subhalos remain dark, lacking baryonic matter \citep{springel2005cosmological, 2019Galax...7...81Z}. 

Gravitational instability also shapes the structure on larger scales, producing the intricate cosmic web \citep{Doroshkevich1970, ZelDovich1970, Shandarin1989}. This large-scale geometry forms a continuous network of dense clusters, elongated filaments, extended sheets, and vast under-dense voids \citep{Bond1996, Cautun2014, Libeskind2018}. The cosmic web is a striking manifestation of nonlinear gravitational evolution, influencing the growth, dynamics, and internal properties of individual haloes \citep{2007ApJ...655L...5A, 2007MNRAS.381...41H, Metuki2015, Alonso2015, GVeena2018, Hellwing2021, GVeena2021}.

High-resolution cosmological simulations have shown that DM haloes are far from smooth, hosting a rich population of bound subhalos \citep{1999ApJ...524L..19M, 1999ApJ...522...82K, 2008Natur.454..735D}. These subhalos are remnants of earlier mergers and accretion events that trace the same gravitational processes that built the cosmic web. The subhalo mass function (SHMF) follows an almost featureless power-law extending to masses far below current simulation resolutions \citep{2008MNRAS.391.1685S, 2021MNRAS.506.4210I}. 
This extrapolation is supported by the redshift- and host-mass independence of the SHMF shape, with analytical models suggesting that substructure accounts for a substantial fraction of the total halo mass when extended down to the smallest scales \citep{2017MNRAS.466.4974M, 2018PhRvD..97l3002H}. The smallest subhalos are eventually set by the microphysical properties of the DM particle \citep{10.1111/j.1745-3933.2011.01074.x, 2012ApJ...749...75S}. Subhalos evolve under tidal interactions that strip their outskirts while often leaving intact dense cores \citep{vdBoschOgiya2018, 2021MNRAS.503.4075G}, making them powerful probes of both DM properties and hierarchical structure formation.

The clumpy substructure of DM haloes also plays a central role in indirect detection, if DM indeed decays or self-annihilates. Since the annihilation rate scales with the square of the density, the dense cores of subhalos can significantly boost the total annihilation luminosity relative to a smooth halo. The combined emission from an entire population of subhalos can therefore enhance the expected annihilation signal, an effect commonly quantified through the subhalo boost factor, i.e. the enhancement relative to the smooth host-halo contribution, \citep{2007PhRvD..75h3526S, 2009NJPh...11j5006B, 2012Prama..79.1021C, 2014MNRAS.442.2271S, 2017MNRAS.466.4974M, 2019Galax...7...68A, 2020MNRAS.492.3662I}, making accurate modeling of the subhalo boost factor crucial for interpreting indirect detection data and constraining particle DM models.

The cosmic web environment strongly shapes halo and subhalo properties. The large-scale environment influences halo formation histories, concentrations, shapes, and spins, reflecting the anisotropic nature of the formation of cosmic structure \citep{2007MNRAS.375..489H, Metuki2015, Metuki2016, Hellwing2021}. Subhalos within dense filaments tend to be more numerous, more concentrated, and exhibit modified radial distributions compared to those in walls or voids \citep{Jiang2016, Ogiya2021, Errani2021, 2025A&A...700A..65M}. In \citep[][hereafter \citetalias{2025A&A...700A..65M}]{2025A&A...700A..65M}, we systematically studied how the cosmic web influences subhalo abundance and internal properties, demonstrating clear, mass-dependent variations in subhalo abundance and internal structure across filaments, walls, and voids, precisely the ingredients that enter the boost-factor modeling developed here.

Despite these advances, most of the existing models for subhalo boost factors are calibrated on simulations of isolated host haloes. These models often assume that the properties of the subhalo depend solely on the mass of the host halo and the radial distance from the halo centre \citep{2014MNRAS.442.2271S, 2017MNRAS.466.4974M, 10.1093/mnras/stac2930}, without explicitly accounting for the influence of the large-scale cosmic web environment. This omission may introduce systematic biases into boost-factor predictions, potentially misestimating the expected annihilation signals from haloes of similar mass but differing formation histories and environments.

In this paper, we investigate how the cosmic web influences subhalo populations and, in turn, DM annihilation signals, using high-resolution cosmological \Nbody~simulations \citep{Bose2016, Hellwing2016}. We examine subhalos residing in different large-scale environments, including filaments, walls, 
and voids, and evaluate their contribution to the total annihilation luminosity. Our analysis employs a new implementation of the multi-scale cosmic web classifier \nexus \citep{Cautun2013} within the \cactus\ framework, enabling robust identification of large-scale structures and a detailed characterization of environment-dependent subhalo properties. The ultimate goal is to develop a physically motivated description of the subhalo boost factor that explicitly accounts for the influence of the large-scale environment, thereby improving predictions for the spatial distribution and amplitude of DM annihilation signals in indirect-detection studies.

The framework of \citet{2014MNRAS.442.2271S}, hereafter \citetalias{2014MNRAS.442.2271S}, provides an environment-independent description of substructure down to the smallest scales, while subsequent work has extended this picture by incorporating the subhalo position within the host and the effects of tidal evolution \citep[e.g.][hereafter \citetalias{2017MNRAS.466.4974M}]{2017MNRAS.466.4974M}. These approaches capture important internal and radial trends, but do not account for systematic variations driven by the large-scale cosmic web.

In this work, we build on existing semi-analytic descriptions by explicitly accounting for the role of the large-scale cosmic web. Our primary quantity of interest is the subhalo boost factor at fixed host-halo mass, expressed relative to the cosmic-mean prediction. Using environment-resolved measurements from simulations, we incorporate the environmental dependence of host-halo concentration, subhalo abundance, and subhalo internal structure into the \citetalias{2014MNRAS.442.2271S} and \citetalias{2017MNRAS.466.4974M} frameworks. The resulting annihilation luminosities and boost factors are therefore model predictions, rather than direct measurements of the full annihilation signal from the simulation. This allows us to quantify coherent, environment-conditioned variations in annihilation luminosities and boost factors that are not captured by models based on host mass and subhalo radius alone.

The paper is organized as follows: in Section~\ref{sec:simulation}, we describe the simulation setup and the methods used to identify haloes and subhalos. Section~\ref{sec:cosmicweb} details our methodology for classifying the cosmic web. In Section~\ref{sec:results}, we present our main results on how subhalo properties and annihilation boost factors depend on the cosmic web environment. Finally, in Section~\ref{sec:summary}, we summarize our findings and discuss their implications for indirect detection of DM. 
\section{\label{sec:simulation} \textsc{\Nbody} Simulations}

Our study uses the simulation of the COpernicus complexio LOw Resolution (COLOR) as the parent volume for the high-resolution COpernicus COmplexio (COCO) DM–only zoom-in \Nbody~simulation \citep{Bose2016, Hellwing2016}. COLOR evolves $1620^3 \approx 4.25 \times 10^9$ particles from an initial redshift of $z_{\rm ini} = 127$ to the present day ($z=0$) in a periodic cubic box of side $70.4~\Mpch$, corresponding to a comoving volume of $3.5 \times 10^{5}~h^{-3}~\mathrm{Mpc^3}$. Each DM particle has a mass of $m_{\rm p} = 6.19 \times 10^6~\Msun$. The simulation assumes a flat $\Lambda$CDM cosmology consistent with the WMAP7 results \citep{2011ApJS..192...18K}, with parameters $\Omega_{\rm m} = 0.272$, $\Omega_{\rm b} = 0.04455$, $\Omega_\Lambda = 0.728$, $h = 0.704$, $n_{\rm s} = 0.967$ and $\sigma_8 = 0.81$.

Dark matter haloes are identified using a Friends-of-Friends (FOF) algorithm with a linking length equal to 0.2 times the mean inter-particle separation \citep{Davis1985}. To ensure adequate resolution for substructure studies, we retain only FOF groups with at least 1000 particles ($M_{\rm FOF} \gtrsim 10^9~\Msun$). The virial mass, \mvir, is defined as the mass enclosed within the virial radius, \rvir, where the mean density is 200 times the critical density at the given redshift, $\rho_{\rm c}(z) = 3H(z)^2/(8\pi G)$. Unless otherwise specified, \mvir{} is used as the definition of the primary halo mass throughout this work.

Subhalos are identified within each FOF halo using the \textsc{SubFind} algorithm \citep{2001MNRAS.328..726S}, which locates over-dense gravitationally bound regions. The subhalo mass corresponds to the total mass of the bound particles. To reduce spurious detections, we consider only subhalos containing at least 100 particles and located within the virial radius of their host halo.

In this work, we use the \textsc{COLOR} halo and subhalo catalog. 
The cosmic web environment is assigned according to the web classification (see Section~\ref{sec:cosmicweb}) at the spatial position of the host-halo center. All subhalos associated with a given host therefore inherit the same large-scale environment label as their parent halo.

\section{\label{sec:cosmicweb} Cosmic Web Classification}

Defining the boundaries of the cosmic web, which consists of voids, walls, filaments, and nodes, is not straightforward. Unlike DM haloes, whose limits can be linked to virialisation, the cosmic web is hierarchical, anisotropic, and multi-scale, so its segmentation relies on algorithmic criteria. Various approaches have been proposed, including methods based on the eigenvalues of the Hessian of the density field \citep{2007A&A...474..315A, 2007IAUS..235..127N, 10.1111/j.1365-2966.2010.17307.x}, the velocity shear tensor \citep{10.1111/j.1365-2966.2012.21553.x, 2017ApJ...845...55P}, combinations of density field and velocity shear information \citep{Cautun2013}, the tidal or deformation tensor \citep{10.1111/j.1365-2966.2009.14885.x}, watershed segmentation of the density field \citep{2010ApJ...723..364A}, cosmic web skeleton construction based on Morse theory \citep{10.1111/j.1365-2966.2011.18394.x}, and the identification of caustics \citep{2018JCAP...05..027F}. A thorough review of these methods is given in \cite{Libeskind2018}.

In this study, we adopt the approach of \citet{Hellwing2021} and employ the \nexus~framework, implemented within our \cactus~algorithm introduced in \citetalias{2025A&A...700A..65M}. The \cactus~package provides an optimized implementation of the multi-scale \nexus~algorithm \citep{Cautun2013}, enabling efficient segmentation of the cosmic web across a wide dynamic range. 
The code, along with detailed documentation, will be made publicly available in Naidoo et al.~(in prep.)\footnote{\cactus is currently being prepared for public release and is not yet available at the time of writing.}

The density field is first computed from the \textsc{COLOR} simulation using a Cloud-In-Cell interpolation onto a $256^3$ grid, yielding a spatial resolution of $0.275~\Mpch$. The dimensionless density field is defined as $f(\vec{x}) = \rho(\vec{x}) / \langle \rho \rangle$ and smoothed over multiple scales with Gaussian filters defined as $R_n = 2^{n/2} R_0$, where $R_0 = 0.5~\Mpch$, producing seven filter scales ranging from $0.5$ to $4~\Mpch$. Filaments and walls are identified using the smoothed logarithmic density field, emphasizing their elongated and planar structures, while nodes are traced with the standard Gaussian-smoothed density to maintain their compact, nearly spherical shape \citep[see also][]{Cautun2014, Cautun2015}.

For each smoothing scale, the normalized Hessian matrix of the smoothed density field is computed as
\begin{equation}
H_{ij, R_n}(\vec{x}) = R_n^2 \, \frac{\partial^2 f_{R_n}(\vec{x})}{\partial x_i \, \partial x_j},
\end{equation}
where the factor $R_n^2$ ensures scale-invariant normalization. The eigenvalues $\lambda_1 \leq \lambda_2 \leq \lambda_3$ are obtained by solving the characteristic equation $\det(H_{ij, R_n} - \lambda I) = 0$, and their relative values and signs are used to determine the morphological signature of each environment. In general, node-like regions correspond to $\lambda_1 \approx \lambda_2 \approx \lambda_3 < 0$, filaments to $\lambda_1 \approx \lambda_2 < 0$ with $|\lambda_3| \ll |\lambda_1|$, and walls to $\lambda_1 < 0$ with $|\lambda_1| \gg |\lambda_2| \approx |\lambda_3|$. Regions not matching any of these criteria are classified as voids. The environmental signature strength $S_{R_n}(\vec{x})$ is computed at each grid cell following the formalism of \citet{Cautun2013}.

The results from all smoothing scales are then combined into a single, scale-independent morphological map by taking the maximum response at each location, $S(\vec{x}) = \max[S_{R_n}(\vec{x})]$, capturing the dominant structural signal independently of filter scale. A hierarchical classification is finally applied, starting with nodes followed by filaments, walls, and voids. Nodes are selected based on mass, density, and signature thresholds, filaments are identified using the corresponding signature threshold and volume criteria, and walls are assigned from the remaining unclassified regions. This approach produces a self-consistent, multi-scale segmentation of the cosmic web that is robust to resolution and noise, while capturing its hierarchical and anisotropic nature. Nodes are included in the classification for completeness, but are not used in our analysis because they occupy a very small volume fraction and are dominated by massive haloes, which have minimal impact on the environment-dependent subhalo ratios.

The specific configuration of the \cactus~algorithm used in this work is detailed in Table~\ref{tab:cosmic_web_params}. These parameters, including grid resolution and smoothing scales, are chosen to ensure a robust multi-scale extraction of the cosmic web components while maintaining consistency with established frameworks \citep[e.g.,][]{Cautun2013, Cautun2014}.
\begin{table*}[ht!]
\centering
\renewcommand{\arraystretch}{1.5}
\setlength{\tabcolsep}{8pt}  
\caption{Numerical parameters and configuration for the \cactus~cosmic web classification.}
\label{tab:cosmic_web_params}
\begin{tabular}{ll} 
\toprule
Parameter             & Value / Method \\ \midrule
Grid Resolution      & $256^3$ ($0.275$~$h^{-1}$\,Mpc cell size) \\
Interpolation Scheme & Cloud-In-Cell (CIC) \\
Smoothing Kernel     & Gaussian, $R_n = 2^{n/2} R_0$ \\
Filter Scales ($R_n$) & $\{0.5, 0.7, 1.0, 1.4, 2.0, 2.8, 4.0\}$~$h^{-1}$\,Mpc \\
Density Transform    & Logarithmic (Filaments/Walls), Linear (Nodes) \\
\bottomrule
\end{tabular}
\end{table*}
\section{Results} \label{sec:results}
In this section, we explore the influence of the large-scale cosmic web environment on DM haloes, their subhalo properties, and the resulting annihilation signals. In particular, we focus on host-halo concentrations, subhalo abundances, and subhalo internal structure and assess how these shape the environment-dependent DM annihilation boost factor. Rather than measuring boosts directly from particle-level $\rho^2$ in individual haloes, we compute them by propagating simulation-derived environmental corrections through the \citetalias{2014MNRAS.442.2271S} and \citetalias{2017MNRAS.466.4974M} semi-analytic frameworks. In practice, the host-halo concentration, the SHMF, and the internal structure of subhalos are modified according to the cosmic web environment and then propagated into the boost calculation. The resulting boost factors are therefore model predictions calibrated by simulation-based environmental trends, rather than direct measurements of the full annihilation signal from the simulation.  

\subsection{Subhalo Boost Factor}
The DM annihilation signal is influenced by contributions from substructure within haloes. This effect is quantified by the boost factor $B(M)$, defined through the total luminosity of a halo including subhalos as $L_{\rm tot} = L_{\rm host} (1 + B(M))$, where $L_{\rm host}$ represents the luminosity of the smooth host component \citep{2008ApJ...686..262K, 2014MNRAS.442.2271S, 2017MNRAS.466.4974M, 2019Galax...7...65S}. Subhalos thus elevate the annihilation signal relative to a smooth DM distribution. 

As a background of our analysis, we first examine the standard approach developed by \citetalias{2014MNRAS.442.2271S}. A key feature of their model is its physically motivated treatment of the concentration--mass relation, $c(M)$. Simple power-law extrapolations of the concentration--mass relation toward low masses, by contrast, can assign excessively high concentrations to the smallest halos.
This leads to a significant overestimation of the boost factor, because the annihilation luminosity scales as $L \propto c^3$. The \citetalias{2014MNRAS.442.2271S} model instead adopts the relation proposed by \citet{2012MNRAS.423.3018P}. 
In that relation, halo concentrations flatten at low masses. Haloes below a certain mass acquire similar values of $c(M)$ that reflect their formation epochs rather than a continued steep power-law rise toward lower masses. This ensures that small haloes collapsing at similar epochs have comparable initial concentrations. As a result, the \citetalias{2014MNRAS.442.2271S} approach provides a more realistic boost prediction than simple power-law extrapolations of $c(M)$ toward low masses.

The \citetalias{2014MNRAS.442.2271S} model recursively accounts for the full hierarchy of substructure via the integral:
\begin{equation}
B(M) = \frac{1}{L(M)} \int_{M_{\mathrm{min}}}^{M} \frac{dN}{dm} \left[1 + B(m)\right] L(m) \, dm,
\label{eq:scp14_boost}
\end{equation}
where $L(M) = 4\pi Mc^{3} / f(c)^2$ (with $f(c) = \ln(1+c) - c/(1+c)$) is the annihilation luminosity of a smooth 
host halo of mass $M$ and concentration $c$. The SHMF is parameterized as $dN/dm = (A/M) (m/M)^{-\alpha}$, with a slope $\alpha = 2$ and normalization $A = 0.012$ to match the $\sim 10\%$ mass fraction in substructure found in high-resolution simulations \cite{2014MNRAS.442.2271S}. The term $B(m)$ represents the boost factor of a subhalo of mass $m$ due to the next level of substructure, and the integration extends down to a minimum subhalo mass $M_{\mathrm{min}}=10^{-6}~M_{\odot}$, which is set by particle physics considerations. We adopt Eq.~\ref{eq:scp14_boost} as the basis for computing the boost factor in our analysis. We then extend this framework by replacing the universal host-halo concentration–mass relation, SHMF, and subhalo structural parameters with environment-dependent values derived from our simulation, enabling the computation of \(B(M, \mathrm{env})\) for haloes in different cosmic web environments. 

In addition to the \citetalias{2014MNRAS.442.2271S} framework, we also consider the distance-dependent subhalo concentration model of \citetalias{2017MNRAS.466.4974M},
which explicitly accounts for both subhalo mass and radial position within the host halo. In this model, the boost factor is computed as
\begin{equation}
\begin{aligned}
B(M) &= \frac{3}{L_{\rm smooth}(M)} 
\int_{M_{\rm min}}^{M} \frac{dN(m)}{dm} \, dm \int_{0}^{1} dx_{\rm sub} \, \\
&\quad \times \big[ 1 + B(m) \big] \, L(m, x_{\rm sub}) \, x_{\rm sub}^2,
\end{aligned}
\label{eq:moline_boost_standard}
\end{equation}
where $L(m, x_{\rm sub})$ is the annihilation luminosity of a subhalo of mass $m$ at a normalized radial coordinate $x_{\rm sub} = R_{\rm sub}/R_{200}$. The integral over $x_{\rm sub}$ accounts for contributions from both inner, high-density and outer, lower-density regions, with $x_{\rm sub}^2$ representing the spherical volume element. $L_{\rm smooth}(M)$ is the luminosity of the smooth host halo. This formulation captures the effects of tidal evolution on subhalo concentrations without yet including explicit environment-dependent modifications.

\subsection{Host halo concentration}
The concentration of a DM halo quantifies how tightly mass is packed toward its center. It is commonly defined as $c \equiv R_\mathrm{vir}/r_s$, where $R_\mathrm{vir}$ is the virial radius and $r_s$ is the scale radius of the halo's density profile. Higher concentrations correspond to haloes that assembled earlier and accumulated mass more efficiently in their central regions. Concentration, therefore, traces both halo assembly history and the larger-scale environment in which a halo evolves.
Haloes in dense regions of the cosmic web, such as filaments, generally assemble earlier and develop higher central densities than those in under-dense regions, consistent with the accelerated growth expected in such environments \citep{Hellwing2021}. This environmental dependence is captured in the concentration--mass relation, which exhibits a systematic variation across the cosmic web \citep{2005ApJ...634...51A, Hellwing2021}. Although the global relation is well described by a smoothly broken power law with a characteristic transition mass, the slope steepens in over-dense regions and flattens at low masses \citep{2014MNRAS.442.2271S, Hellwing2016}. In practice, filament haloes tend to be more concentrated than the cosmic mean, while void haloes consistently display the lowest concentrations across the relevant mass range.

In our analysis, we adopt the concentration--mass relation from the \citetalias{2014MNRAS.442.2271S} model as our baseline, based on the physically motivated parametrization of \citet{2012MNRAS.423.3018P} for the median halo concentration, $c(M)$, which naturally captures the flattening of concentrations at low masses.
We adopt the specific $z=0$ fitting function provided by \citetalias{2014MNRAS.442.2271S}, 
using the fitting parameters from \citet{Hellwing2016}, derived for the combined COCO+COLOR simulation suite employed in this work.
As noted by Ref. \cite{Hellwing2016}, the original \citetalias{2014MNRAS.442.2271S} parametrization systematically over-predicts the concentrations of haloes with $\mvir \lesssim 10^{11}\Msun$,
due to a mismatch in the low-mass slope. To remedy this, \citet{Hellwing2016} refitted the same functional form using the combined \textsc{COCO}+\textsc{COLOR} simulation suite. The resulting parametrization accurately reproduces the median concentration to better than 1\% through the entire halo mass range $10^{-6}$ to $10^{15}\Msun$.

\begin{figure*}[!htbp]
  \centering
  \includegraphics[width=\linewidth]{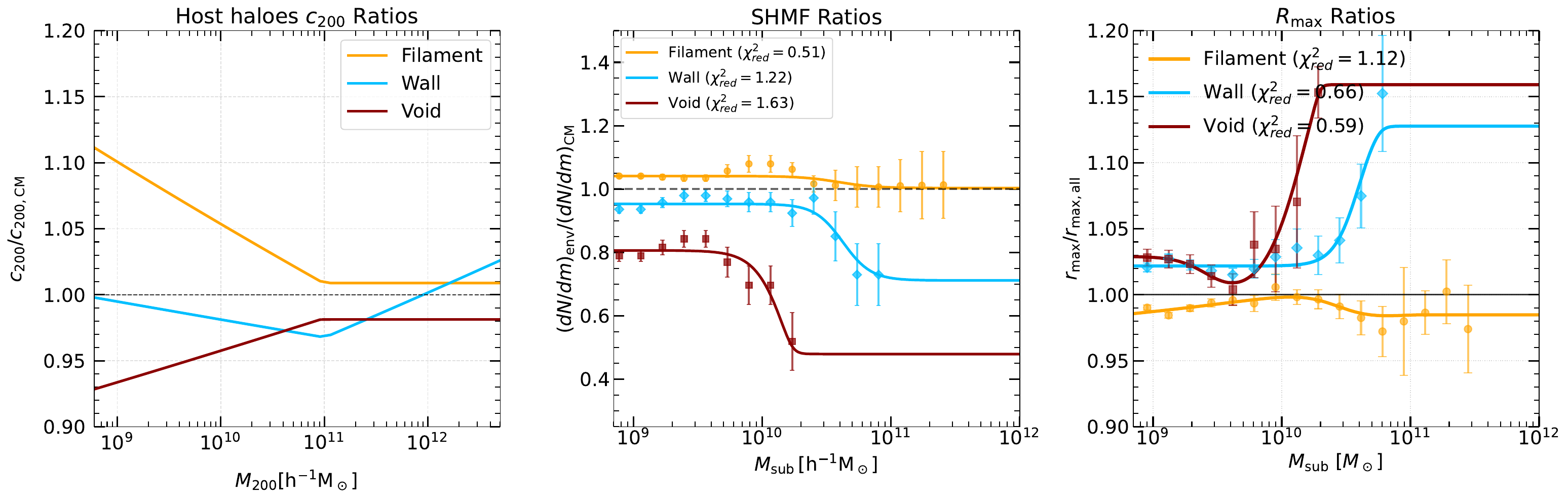}
  \caption{\textit{Left Panel:} Host halo concentration relative to the cosmic mean as a function of halo mass. The curves show median concentration ratios for different cosmic web environments, taken from \citetalias{Hellwing2021} using the COCO+COLOR simulations (Table I, therein). Together with the middle and right panels, this panel summarizes the three environment-dependent ingredients propagated into the boost calculation.
  \textit{Middle Panel:} Differential SHMF ratios, $(dN/dm)_{\rm env}/(dN/dm)_{\rm CM}$, where the data points are calculated from simulation data for each environment. Solid lines show the corresponding single power-law transition fits to these differential ratios (Eq.~\ref{eq:single_sigmoid}).
  These fitted differential curves are used in Eq.~\ref{eq:shmf_env_diff} for all subsequent calculations. \textit{Right Panel:} Subhalo $\rmax$ ratio relative to the cosmic mean as a function of subhalo mass $m$. Solid curves show the best-fit power-law transition models: a single power-law transition form is used for the wall environment, while a double power-law transition model with a controlled high-mass plateau is adopted for filaments and voids (Eqs.~\ref{eq:single_sigmoid} and \ref{eq:double_sigmoid}). 
  }
  \label{fig:ratio_plots}
\end{figure*}

To account for environmental effects, we apply multiplicative ratios from \citet{Hellwing2021}, hereafter \citetalias{Hellwing2021}, to the baseline \citetalias{2014MNRAS.442.2271S} concentration--mass relation,
\begin{equation}
c(\mvir, \mathrm{env}) = c_{\mathrm{SCP14}}(\mvir) \times \mathcal{R}_c(\mathrm{env}, \mvir),
\label{eq:host_conc_env}
\end{equation}
Here, $\mathcal{R}_c(\mathrm{env}, \mvir)$ denotes the ratio of the median halo concentration in a given environment to the cosmic mean at the same mass, as illustrated in the left panel of Fig.~\ref{fig:ratio_plots}. These environment-dependent concentration ratios are taken directly from the measurements of \citetalias{Hellwing2021}, and were derived using the combined \textsc{COCO}+\textsc{COLOR} simulation suite.
Filament haloes are more concentrated than the cosmic mean at low masses, void haloes are less concentrated, and wall haloes remain closer to the cosmic mean.
This multiplicative scaling preserves the mass-dependent shape of the \citetalias{2014MNRAS.442.2271S} relation while explicitly embedding the empirically observed environmental variation. The ratios for filaments start above unity at low masses and decrease toward the cosmic mean, leveling off beyond $10^{11}\Msun$. The ratios for voids begin below unity and increase toward the cosmic mean, also remaining nearly constant past $10^{11}\Msun$. For walls, the ratios start near the cosmic mean, decrease slightly, and then rise at higher masses.
These environment-dependent concentrations, $c(\mvir, \mathrm{env})$, are subsequently used in the calculation of the boost factor, entering Eq.~\ref{eq:scp14_boost} as the host halo concentration to account for environmental modulation of the host density profile.
\subsection{Subhalo mass function}
The subhalo mass function characterizes the abundance of subhalos within a host halo as a function of subhalo mass, $m$. It provides a statistical description of the hierarchical substructure across a wide mass range. The abundance and mass distribution of subhalos are closely linked to the host halo's formation history and dynamical evolution, which are influenced by the large-scale cosmic web environment.

Analysis of the high-resolution \textsc{COLOR} simulations in \citetalias{2025A&A...700A..65M} showed that subhalo populations depend systematically on environment. Halos in filaments host a larger number of bound substructures than those in walls or voids, reflecting earlier formation times and higher survival rates of substructures in dense regions. Conversely, haloes in under-dense regions form later and retain fewer subhalos, whereas haloes in walls exhibit abundances close to the cosmic mean.

In \citetalias{2025A&A...700A..65M}, we quantified the environmental dependence through cumulative SHMF ratios. These cumulative measurements provide the baseline environmental trends that we adopt here. For the boost-factor calculation, however, we require differential SHMF ratios, which we compute directly from the simulation data and use in Eq.~\ref{eq:scp14_boost}:
\begin{equation}
\mathcal{R}_{\rm SHMF}(m) = \frac{dN/dm_{\rm env}}{dN/dm_{\rm CM}},
\end{equation}
where $dN/dm_{\rm env}$ and $dN/dm_{\rm CM}$ are the differential number of subhalos per unit mass in a given environment and in the cosmic mean, respectively. 

To describe the mass dependence of these differential ratios, we fit a power-law transition function of the form
\begin{equation} 
\mathcal{R}_{\rm SHMF}(m) = p + \frac{q - p}{1 + \left(m_b/m\right)^{\nu}} \,. 
\label{eq:single_sigmoid}
\end{equation} 
where $p$ and $q$ are the low- and high-mass plateaus, respectively, $m_b$ sets the transition mass, and $\nu$ controls the steepness of the transition.
Fits are performed over environment-specific mass ranges to ensure robust parameter estimation and avoid high-mass regions with limited statistics. The lower bound is set to $6.2 \times 10^{8}\Msun$, corresponding to 100 times the particle mass, while the upper bound is given by the maximum halo mass, $m_{\rm max}$, in each environment, as listed in Table~\ref{tab:shmf_sigmoid}. 

The best-fit parameters for the subhalos hosted by the haloes in the filaments, walls, and voids are reported in Table~\ref{tab:shmf_sigmoid}. Each environment exhibits a distinct plateau and transition mass, reflecting the influence of large-scale structure on subhalo abundances.
For the filament and void fits, the inferred transition mass lies close to the upper edge of the fitted interval, so it should be interpreted mainly as an effective scale governing the smooth turnover of the interpolation, rather than as a sharply resolved physical break within the sampled mass range.
The middle panel of Fig.~\ref{fig:ratio_plots} shows the measured $\mathcal{R}_{\rm SHMF}(m)$ with error bars representing bootstrap errors on the mean.
The corresponding reduced $\chi^2$ values are also shown as a compact indicator of fit quality. Typical residuals remain at the few-percent level, staying within about $5\%$ over most of the fitted mass range; larger deviations occur only in the more weakly populated bins, where the statistical uncertainties are also largest.

\begin{table*}[!ht]
\centering
\renewcommand{\arraystretch}{1.5}
\setlength{\tabcolsep}{8pt}      
\caption{Best-fit parameters of the power-law transition model for the environment-dependent SHMF ratios \(Eq.~\ref{eq:single_sigmoid}\) The maximum halo mass, $m_\mathrm{max}$, indicates the upper limit of the mass range used for fitting each environment. 
}
\label{tab:shmf_sigmoid}
\begin{tabular}{lccccc}
\hline 
Environment & $p$ & $q$ & $\log_{10} m_b\ [\Msun]$ & $\nu$ & $\log_{10} m_\mathrm{max}[\Msun]$  \\
\hline
Filament & 1.04 $\pm$ 0.01 & 1.00 $\pm$ 0.05 & 10.59 $\pm$ 0.79 & 2.97 $\pm$ 1.04 & 11.00 \\
Wall     & 0.95 $\pm$ 0.01 & 0.71 $\pm$ 0.11 & 10.63 $\pm$ 0.18 & 4.01 $\pm$ 1.93 & 10.78 \\
Void     & 0.81 $\pm$ 0.01 & 0.37 $\pm$ 0.28 & 10.10 $\pm$ 0.16 & 3.49 $\pm$ 1.63 & 10.23 \\
\hline
\end{tabular}
\end{table*}
To incorporate these results in calculations that require the differential SHMF, we employ:
\begin{equation}
\frac{dN}{dm}(m, \mvir, \mathrm{env}) = \frac{dN_{\mathrm{SCP14}}}{dm}(m, \mvir) \times \mathcal{R}_{\rm SHMF}(\mathrm{env}, m),
\label{eq:shmf_env_diff}
\end{equation}
where we use the fitted differential ratios $\mathcal{R}_{\rm SHMF}(m)$. These fitted ratios provide a smooth working representation of the environmental correction and remain consistent with the finite-difference estimates from the simulation data. This ensures that the differential SHMF applied in Eq.~(\ref{eq:shmf_env_diff}) accurately reflects the environmental trends measured in \citetalias{2025A&A...700A..65M}.

By applying  $\mathcal{R}_{\mathrm{SHMF}}(\mathrm{env}, m)$ in the computation of the boost factor (Eq.~\ref{eq:scp14_boost}), the subhalo abundance at each mass is rescaled to include these environmental variations. This ensures that the recursive evaluation of the annihilation luminosity captures both the hierarchical structure of the host halo and the modulation imposed by the large-scale cosmic web, which can substantially affect the predicted DM annihilation signal.

\subsection{Subhalo concentration}
The internal density structure of subhalos strongly influences their contribution to the DM annihilation signal. We characterize subhalo compactness using the characteristic concentration, $c_v$, defined as \citep{2007ApJ...667..859D, 2008Natur.454..735D, 2008MNRAS.391.1685S}:

\begin{equation}
c_v = 2\left(\frac{\vmax}{H_0 \rmax}\right)^2,
\label{eq:cv}
\end{equation}
where $\vmax$ is the maximum circular velocity, $\rmax$ the radius at which it occurs, and $H_0$ the Hubble constant,
included so that $c_v$ remains a dimensionless concentration proxy.
This definition is particularly suitable for subhalos, whose outer regions are often stripped tidally and lack well-defined virial radii. Unlike traditional mass-dependent $\cvir$, which depends on virial mass and radius, $c_v$ relies on $\vmax$ and $\rmax$, which remain well-resolved in simulations even for stripped subhalos. Therefore, $c_v$ provides a robust, model-independent measure of subhalo inner densities across different environments and host masses.

The environmental dependence of $c_v$ is derived from the ratio of the mean values of $\rmax$ at fixed $\vmax$, using the environment-resolved relations $\vmax$--$\rmax$ of \citetalias{2025A&A...700A..65M}. For a given $\vmax$, a smaller $\rmax$ corresponds to a more compact subhalo of higher density, while a larger $\rmax$ indicates a more extended subhalo of lower density. The subhalos in the filaments exhibit a smaller $\rmax$ at fixed $\vmax$, reflecting earlier formation, while those in the voids have a larger $\rmax$, indicative of lower densities.

At fixed $\vmax$ Eq.~\ref{eq:cv} implies $c_v \propto (\vmax / \rmax)^2$, and therefore $c_v \propto R_{\max}^{-2}$ at fixed $V_{\max}$,
Let $R_{\max,\rm env}$ and $R_{\max,\rm CM}$ denote the environment-specific and cosmic-mean values of $\rmax$, respectively. The concentration ratio can then be written as
\begin{equation}
\begin{aligned}
\frac{c_{\rm sub,{env}}}{c_{\rm sub,CM}}
&= \frac{c_{v,\rm{env}}}{c_{v,\rm{CM}}} 
= \left(\frac{R_{\max,\rm CM}}{R_{\max,\rm env}}\right)^2 
 = \mathcal{R}^{-2}_{\rm max, sub}(\mathrm{env}, m).
\end{aligned}
\end{equation}

We use $\vmax$ as the conditioning variable because it traces the depth of the inner subhalo potential and is comparatively insensitive to tidal mass loss, while $\rmax$ captures the scale at which the circular velocity peaks. At fixed $\vmax$, variations in $\rmax$ therefore encode changes in compactness relevant for the annihilation luminosity and naturally lead to the inverse-square correction below. Alternative conditionings, such as measuring the environmental variation of $\vmax$ at fixed $\rmax$, probe the same $\vmax$--$\rmax$ relation from a different direction and can change the detailed amplitude of the correction; we retain the fixed-$\vmax$ calibration as the fiducial choice because it is directly tied to the profile-independent $c_v$ proxy used for tidally stripped subhaloes.

In the \citetalias{2014MNRAS.442.2271S} framework, subhalo concentrations, $c_{\mathrm{SCP14, sub}}$, are assumed to follow the concentrations of field haloes of the same mass. Low-mass haloes collapse nearly simultaneously in the early Universe, producing similar natal concentrations, which results in a flattening of $c(M)$ at small masses. Field-halo concentrations thus provide a reasonable approximation for subhalos.

To account for environmental modulation, we scale the \citetalias{2014MNRAS.442.2271S} subhalo concentrations using the inverse-square of the $\rmax$ ratio:
\begin{equation} 
c_{\mathrm{sub}}(m, \mathrm{env}) = c_{\mathrm{SCP14,sub}}(m) \times \mathcal{R}^{-2}_{\mathrm{max,sub}}(\mathrm{env}, m), 
\label{eq:csub_env} 
\end{equation} 
The inverse-square scaling ensures that subhalos with smaller $R_{\max}$, indicating higher central densities, receive larger concentrations, while those with larger $R_{\max}$ have reduced $c_{\rm sub}$. This preserves physical consistency while capturing the systematic variations in the internal structure of subhalos across different cosmic web environments.

For walls, the mass dependence of $\mathcal{R}_{\mathrm{max, sub}}$ is modelled using the power-law transition function (Eq.~\ref{eq:single_sigmoid}), with environment-specific parameters listed in Table~\ref{tab:rmax_sigmoid_combined}. For voids and filaments, where the trend of the $\rmax$ ratios is more complex, we adopt a power-law function of two transitions with a controlled high-mass plateau: 

\begin{equation}
\mathcal{R}_{\mathrm{max,sub}}(m) =
\begin{cases}
\displaystyle 
b + \frac{c - b}{1 + \left(m_{t1}/m\right)^{s_1}} \\[1mm]
\quad + \frac{d - c}{1 + \left(m_{t2}/m\right)^{s_2}}, & m \le m_\mathrm{pl} \\[1mm]
\mathcal{R}_{\mathrm{max,sub}}(m_\mathrm{pl}), & m > m_\mathrm{pl}
\end{cases}
\label{eq:double_sigmoid}
\end{equation}
where $b$, $c$, and $d$ are the low-mass, intermediate, and high-mass plateaus, respectively. Parameters $m_{t1}$ and $m_{t2}$ are the transition masses, and $s_1$ and $s_2$ control the steepness of the respective transitions. The plateau mass $m_\mathrm{pl}$ above which the fit is kept constant is determined by the last data value, denoted as $m_{\mathrm{max}}$ in Table~\ref{tab:rmax_sigmoid_combined},
to prevent the high-mass tail from diverging due to limited statistics. This controlled plateau ensures that the two-transition power-law function accurately captures the valley and overall trend of the $\rmax$ ratios in voids and filaments.

Table~\ref{tab:rmax_sigmoid_combined} lists the best-fit parameters for the environment-dependent $\rmax$ ratios. 
The fit reproduces the measured $\mathcal{R}_{\mathrm{max,sub}}$ ratios to within about $1$--$2\%$ over most of the fitted range, with somewhat larger deviations appearing only in the highest-mass bins where the uncertainties increase substantially.
These ratios indicate that the subhalos in filaments have a smaller $\rmax$ relative to the cosmic mean, corresponding to higher characteristic densities, while the subhalos in walls display intermediate values, and those in the voids exhibit the largest $\rmax$, reflecting lower internal densities. Across the subhalo mass range, these differences illustrate that the internal structure of subhalos is modulated by their location within the cosmic web.

The mass-dependent behavior of the $\rmax$ ratio captures the systematic variation in subhalo compactness. Subhalos in denser environments are more centrally concentrated, whereas those in under-dense regions are comparatively less concentrated. This modulation affects the expected DM annihilation signal, since subhalos with smaller $\rmax$ produce stronger contributions to the luminosity. The right panel of Fig.~\ref{fig:ratio_plots} illustrates these trends, showing the differences in characteristic subhalo density across filaments, walls, and voids. Consequently, the $\rmax$ ratios provide a quantitative measure of how environmental conditions influence subhalo structure.

This environment-dependent characterization of subhalo internal structure provides the necessary input for computing the annihilation boost factor in host haloes, which we describe in the following subsection. 
\begin{table*}[!ht]
\centering
\renewcommand{\arraystretch}{1.7} 
\caption{Best-fit parameters for environment-dependent subhalo $r_{\rm{max}}$ ratios using fitted transition models.}
\resizebox{\textwidth}{!}{
\begin{tabular}{l c c c c c c c c}
\hline
Environment & $b$ & $c$ & $\log_{10} m_{t1} [\Msun]$ & $s_1$ & $d$ & $\log_{10} m_{t2} [\Msun]$ & $s_2$ & $\log_{10} m_\mathrm{max} [\Msun]$ \\
\hline
Filament & 0.97 $\pm$ 0.20 & 1.03 $\pm$ 0.85 & 10.00 $\pm$ 0.78 & 0.42 $\pm$ 0.22 & 1.00 $\pm$ 0.68 & 10.43 $\pm$ 0.69 & 3.02 $\pm$ 1.35 & 11.00 \\
Wall     & 1.02 $\pm$ 0.01 & 1.14 $\pm$ 0.12 & 10.60 $\pm$ 0.29 & 3.78 $\pm$ 1.03 & ... & ... & ... & 10.78 \\
Void     & 1.03 $\pm$ 0.01 & 0.99 $\pm$ 0.18 & 9.44 $\pm$ 0.70 & 3.50 $\pm$ 0.95 & 1.25 $\pm$ 1.06 & 10.20 $\pm$ 1.27 & 2.61 $\pm$ 0.45 & 10.23 \\
\hline
\end{tabular}
}
\label{tab:rmax_sigmoid_combined}
\end{table*}
\subsection{Environment-dependent luminosity and boost predictions}
\begin{figure}
    \centering
    \includegraphics[width=\linewidth]{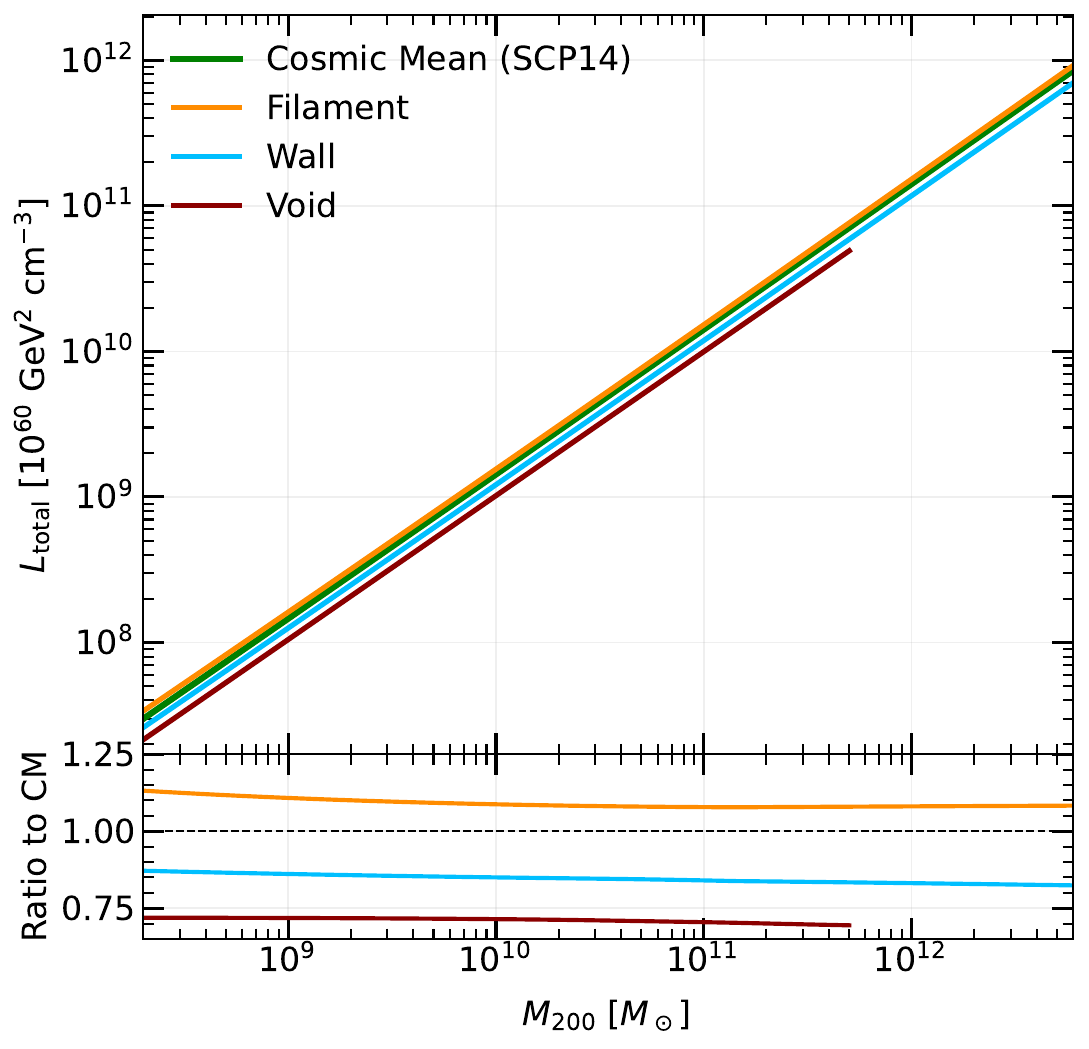}
    \caption{Total halo luminosity as a function of host halo mass \mvir{} for different cosmic environments, including filaments, walls, and voids. The green line shows the cosmic mean prediction from the original \citetalias{2014MNRAS.442.2271S} model without environmental modifications. These curves are not direct $\rho^2$ measurements from simulations but are model predictions obtained by propagating environment-dependent ingredient ratios into the \citetalias{2014MNRAS.442.2271S} framework. The vertical scale is shown in the conventional annihilation-luminosity units of $10^{60}\,\mathrm{GeV}^2\,\mathrm{cm}^{-3}$. The bottom panel shows the ratio of environment-dependent halo luminosities to the cosmic mean. The curves are deterministic from the fits and do not include halo-to-halo scatter.
    \label{fig:halo_luminosity}}
\end{figure}

To assess the impact of the environment on halo properties and their potential DM annihilation signals, we compute the total annihilation luminosity of host haloes, including their substructure. Here, following standard usage in the boost-factor literature, ``annihilation luminosity'' denotes the volume integral of $\rho^2$ up to particle-physics prefactors, rather than radiative power in watts. Fig.~\ref{fig:halo_luminosity} shows the total halo luminosity, $L_{\rm tot}$, as a function of host mass \mvir{} for filaments, walls, and voids, alongside the cosmic mean. Haloes residing in filaments exhibit the highest total luminosities at fixed mass, reflecting both a higher number of subhalos and increased subhalo densities, whereas haloes in voids are the least luminous. Wall haloes follow the cosmic mean trend closely, with minor deviations at intermediate masses. It is important to note that the environmental trend in the total annihilation luminosity need not mirror that of the boost factor. Because the boost is defined relative to the smooth host contribution, environment-dependent changes in host-halo concentration also modify the denominator, so a population with a higher total luminosity can still exhibit a suppressed boost relative to the cosmic mean.

Having quantified how host halo concentration, the SHMF, and subhalo internal density structure vary systematically across the cosmic web, we now incorporate these environmental effects into the computation of the DM annihilation boost factor. Our approach extends the standard \citetalias{2014MNRAS.442.2271S} framework, originally expressed in Eq.~\ref{eq:scp14_boost}, by replacing the universal concentration–mass relation, SHMF, and subhalo structural parameters with their environment-dependent counterparts:
\begin{equation} 
\begin{split}
&B(M, \mathrm{env}) = \frac{1}{L(M, \mathrm{env})} 
\int_{M_{\mathrm{min}}}^{M} \frac{dN}{dm}(m, M, \mathrm{env}) \,\\ 
&\times \big[ 1 + B(m, \mathrm{env}) \big] \, L(m, \mathrm{env}) \, dm, 
\label{eq:env_boost} 
\end{split}
\end{equation}
where $L(M, \mathrm{env})$ denotes the annihilation luminosity of the smooth host halo, calculated using the environment-dependent concentration $c(M, \mathrm{env})$ (\citetalias{Hellwing2021}, Fig. 5 and Table 2 therein). The term $L(m, \mathrm{env})$ accounts for the internal structure of subhalos through their environment-dependent $\rmax$ ratios (Eq.~\ref{eq:csub_env}). By explicitly incorporating variations in both subhalo abundance and internal density, Eq.~\ref{eq:env_boost} generalizes the recursive formulation of \citetalias{2014MNRAS.442.2271S}, capturing the influence of the cosmic web on the DM distribution within host haloes and their substructure. For computational efficiency, we limit the calculation to two hierarchical levels of substructure (subhalos and sub-subhalos). 

In Eq.~\ref{eq:env_boost}, the sub-subhalo boost contribution assigned to each first-level subhalo is computed using the same environment-dependent correction as its parent halo, because no explicit calibration currently exists for environmental effects beyond the first level of substructure. This is a simplifying assumption: nested subhaloes have undergone additional tidal processing and may therefore retain weaker environmental memory than first-level subhaloes. To test its impact, we recomputed the boost using cosmic mean properties for the sub-subhalo contribution. The alternative treatment changes \(B(M,\mathrm{env})/B_{\rm CM}(M)\) by approximately \(5\)--\(9\%\) in filaments and \(7\)--\(8\%\) in walls, rising to \(15\)--\(17\%\) in voids over the mass range considered. It can also modify the relative ordering of the environmental curves at the low-mass end, while at larger host masses the qualitative filament--wall--void hierarchy of the fiducial calculation is recovered. A substantial cosmic-web modulation nevertheless remains, so the central conclusion that the boost depends on large-scale environment is unchanged; however, the treatment of nested substructure should be regarded as a modelling uncertainty in the detailed amplitude and low-mass environmental ranking.

\begin{figure}
    \centering
    \includegraphics[width=\linewidth]{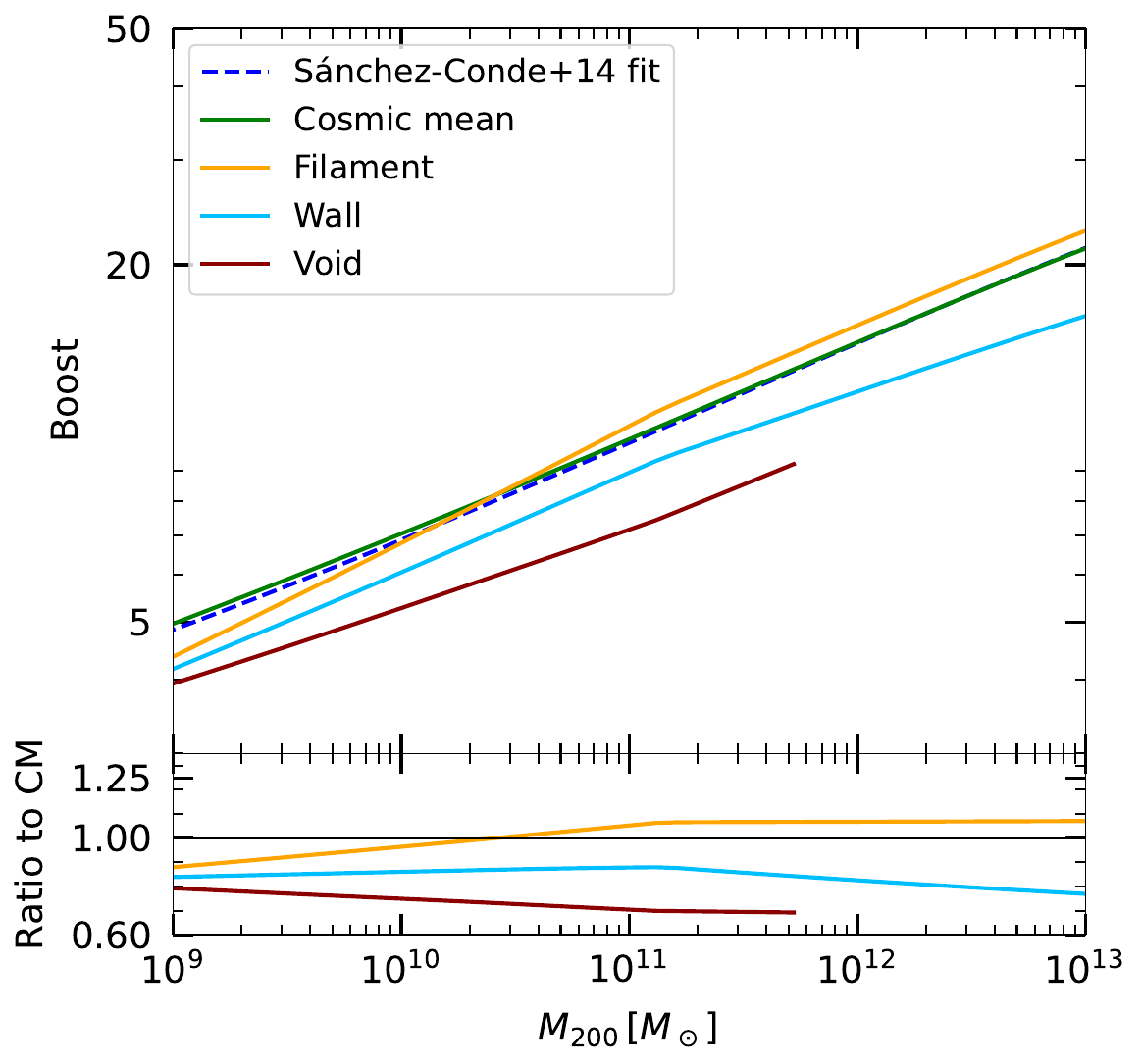}
    \caption{\textit{Upper panel:} 
    Baseline \citetalias{2014MNRAS.442.2271S}-based environment-dependent subhalo boost factors as a function of host halo mass for each cosmic web environment. The solid green line shows the cosmic-mean boost, while the black dashed line represents the fit from \citetalias{2014MNRAS.442.2271S}. The curves shown here are direct model predictions obtained from the fitted environmental ingredients and do not include halo-to-halo scatter. \textit{Lower panel:} Ratio of the environment-specific boost factors to the cosmic mean. The curves are deterministic from the fits and do not include uncertainties from halo-to-halo scatter. 
        \label{fig:Boost_scp}}
\end{figure}

Combining the physical interpretation of subhalo abundance, internal density, and host halo concentration with the quantitative power-law transition fits, Fig.~\ref{fig:Boost_scp} summarises the baseline \citetalias{2014MNRAS.442.2271S}-based environment-dependent boost prediction obtained by propagating the fitted concentration, SHMF, and subhalo-structure corrections into Eq.~\ref{eq:scp14_boost}. The lower panel of Fig.~\ref{fig:Boost_scp} shows the ratio of these environment-dependent boosts to the cosmic-mean prediction. Filament haloes display a smooth increase with host mass, starting at a suppression of $\sim 0.85$ at low masses, crossing unity near $M \sim 10^{11}\Msun$, and reaching an enhancement of $\sim 1.07$ at the high-mass end. Wall haloes reach a peak ratio of approximately $0.90$ at intermediate masses, before decreasing to $\sim 0.76$ at the high-mass end. Void haloes remain suppressed across the entire mass range, with ratios as low as $\sim 0.7$, reflecting their lower subhalo abundance and reduced internal densities. It is worth stressing that the cosmic-mean reference here is the global boost factor prediction, not the average of the three explicitly shown environments. At low host masses, all three displayed environments can therefore lie below the cosmic mean simultaneously. This reflects both the fact that the reference includes the full halo population used to define the cosmic mean, and that the boost is normalised by the smooth host contribution, so environment-dependent changes in host concentration can suppress the boost ratio even when the total annihilation luminosity is not correspondingly reduced.

The results presented here should be interpreted as deterministic model predictions based on best-fit environment-dependent corrections to the adopted semi-analytic framework. In the present implementation, we do not propagate halo-to-halo scatter, uncertainties in the fitted environmental ratios, or classification uncertainties associated with the Cosmic Web assignment. We therefore regard the trends reported here as a first characterization of the systematic impact of large-scale environment on boost-factor modeling, with a full uncertainty propagation left for future work.

\begin{table*}[!ht]
\centering
\renewcommand{\arraystretch}{1.4}
\setlength{\tabcolsep}{5pt}
\caption{Best-fitting values of the polynomial coefficients given in Eq.~\ref{eq:moline_poly} for the fit to the subhalo boost factor \ in the cosmic-mean and environment-dependent cases shown in Fig.~\ref{fig:boost_moline_tidal}.
}
\label{tab:boost_poly_coeffs}
\begin{tabular}{lcccc}
\hline
Environment & $b_0$ & $b_1$ & $b_2$ & $b_3$ \\
\hline
Cosmic mean & $0.468 \pm 0.094$ & $-0.058 \pm 0.030$ & $1.666\times10^{-2} \pm 3.079\times10^{-3}$ & $-4.232\times10^{-4} \pm 1.024\times10^{-4}$ \\
Filament    & $1.385 \pm 0.095$ & $-0.423 \pm 0.030$ & $5.769\times10^{-2} \pm 3.106\times10^{-3}$ & $-1.811\times10^{-3} \pm 1.033\times10^{-4}$ \\
Wall        & $1.068 \pm 0.134$ & $-0.320 \pm 0.044$ & $4.867\times10^{-2} \pm 4.727\times10^{-3}$ & $-1.684\times10^{-3} \pm 1.643\times10^{-4}$ \\
Void        & $-0.430 \pm 0.125$ & $0.215 \pm 0.043$ & $-1.138\times10^{-2} \pm 4.958\times10^{-3}$ & $3.915\times10^{-4} \pm 1.848\times10^{-4}$ \\
\hline
\end{tabular}
\end{table*}
Extending the standard distance-dependent subhalo concentration model of \citetalias{2017MNRAS.466.4974M} (Eq.~\ref{eq:moline_boost_standard}), we now include the effects of the cosmic web environment, moving from the baseline \citetalias{2014MNRAS.442.2271S}-based calculation of Fig.~\ref{fig:Boost_scp} to a fiducial model that also incorporates the radial dependence of subhalo structure within the host halo. In this framework, the subhalo concentration depends on both its mass and radial position within the host halo. The normalized radius is defined as $x_{\rm sub} = R_{\rm sub}/R_{200}$, representing the subhalo’s distance from the host center in units of the host virial radius. The full parametrization of the concentration is provided in Appendix~\ref{app:moline_concentration}. In this fiducial variant, tidal stripping is not introduced as an additional environment-dependent correction, but is incorporated implicitly through the distance-dependent subhalo concentration prescription adopted from \citetalias{2017MNRAS.466.4974M}.
Using these distance- and environment-dependent concentrations, the subhalo boost factor can be expressed as

\begin{equation}
\begin{split}
&B(M, \mathrm{env}) = \frac{3}{L_{\rm sm}(M, \mathrm{env})} 
\int_{M_{\rm min}}^{M} \frac{dN(m, \mathrm{env})}{dm} \, dm \\
& \times \int_{0}^{1} dx_{\rm sub} \, \big[ 1 + B(m, \mathrm{env}) \big] \, L(m, x_{\rm sub},env) \, x_{\rm sub}^2,
\end{split}
\label{eq:moline_boost_env}
\end{equation}
where $L(m, x_{\rm sub}, \mathrm{env})$ denotes the annihilation luminosity of a subhalo of mass $m$ at radius $x_{\rm sub}$, incorporating both distance-dependent concentrations and environmental modulation. The factor $x_{\rm sub}^2$ accounts for the spherical volume element, while $L_{\rm sm}(M, \mathrm{env})$ is the luminosity of the smooth host halo including environmental effects. This formulation generalizes the original model (Eq.~\ref{eq:moline_boost_standard}) by explicitly including cosmic web variations in subhalo abundance and internal structure. Relative to the \citetalias{2014MNRAS.442.2271S}-based result of Fig.~\ref{fig:Boost_scp}, it therefore adds the radial dependence of subhalo concentrations within the host halo while preserving the same environment-dependent ingredient corrections.

\begin{figure}[!htbp]
  \centering
  \includegraphics[width=\linewidth]{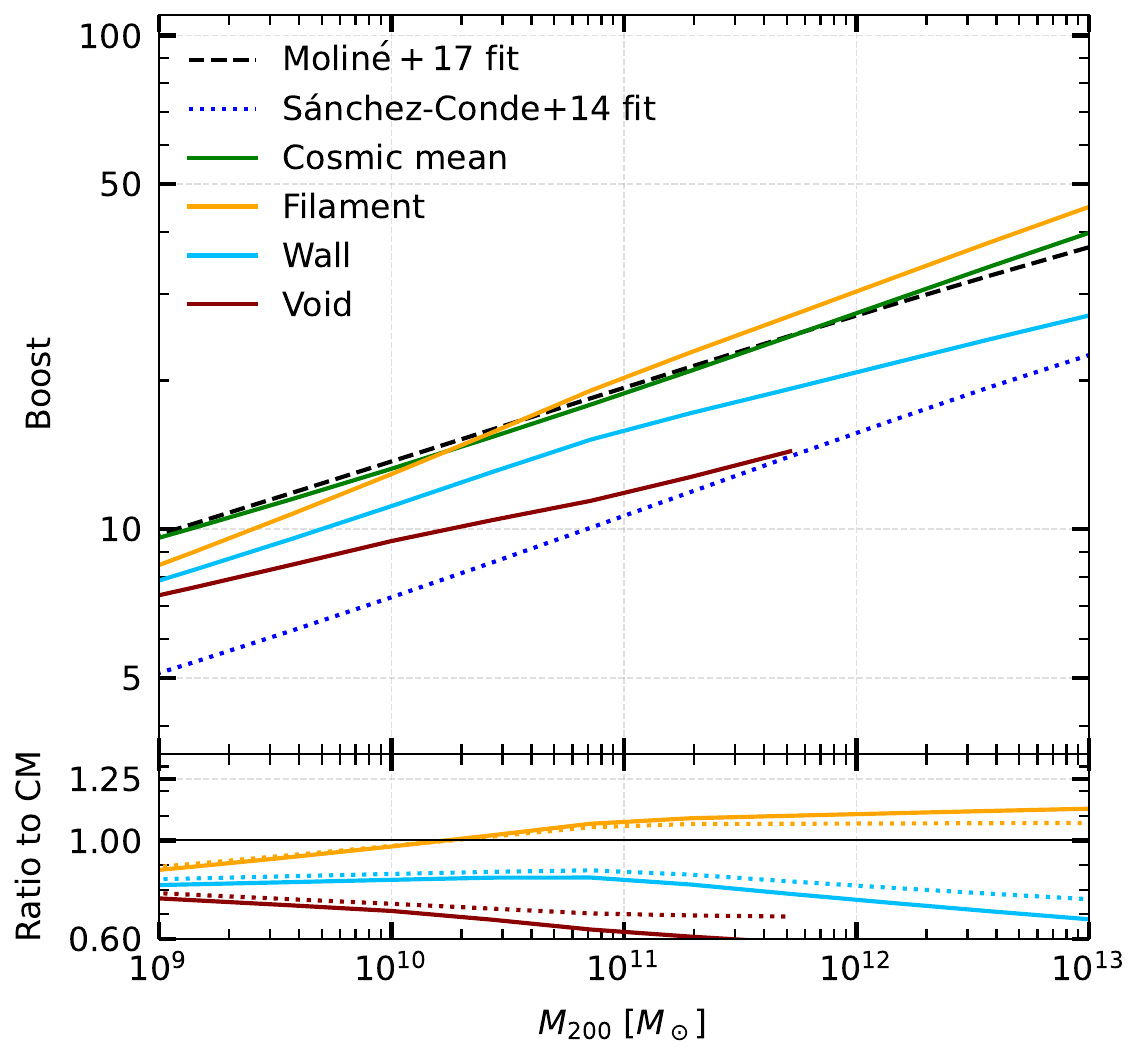}
  \caption{Environment-dependent subhalo boost factor as a function of \mvir{} mass of the host. \textit{Upper panel:} Boost factors for haloes in different cosmic web environments. The green solid line shows the cosmic-mean prediction from \citetalias{2017MNRAS.466.4974M} without environment modification, the blue dotted line shows the \citetalias{2014MNRAS.442.2271S} fit, and the black dashed lines indicate the polynomial fits for each environment. Tidal stripping is included in the calculations. \textit{Lower panel:} Ratio of each environment to the cosmic-mean prediction, showing the relative effect of the cosmic web. Compared to Fig.~\ref{fig:Boost_scp}, these curves are computed in the \citetalias{2017MNRAS.466.4974M}-based distance-dependent framework, in which subhalo concentrations depend explicitly on both subhalo mass and radial position within the host halo.}
  \label{fig:boost_moline_tidal}
\end{figure}

Fig.~\ref{fig:boost_moline_tidal} shows the resulting environment-dependent subhalo boost factors obtained in the fiducial \citetalias{2017MNRAS.466.4974M}-based model, which differs from Fig.~\ref{fig:Boost_scp} by explicitly including the radial dependence of subhalo concentrations and the associated tidal-evolution effects encoded in that prescription. The upper panel presents the boost as a function of host \mvir{} mass for each cosmic web environment. The green solid line indicates the cosmic-mean prediction from \citetalias{2017MNRAS.466.4974M} without environment modifications, the blue dotted line shows the \citetalias{2014MNRAS.442.2271S} fit, and the black dashed lines correspond to polynomial fits introduced below in Eq.~\ref{eq:moline_poly} for each environment. The lower panel displays the ratio to the cosmic-mean prediction, highlighting the relative impact of the cosmic web.

Tidal stripping, as encoded in the adopted distance-dependent \citetalias{2017MNRAS.466.4974M} concentration prescription, primarily removes outer subhalo material while leaving the dense inner regions largely intact. High-resolution simulations have shown that this process reduces subhalo mass and size relative to field haloes \citep{2001ApJ...559..716T, 2004ApJ...609..482K, 2007ApJ...667..859D, 2008Natur.454..735D, 2019MNRAS.485..189O, 2020MNRAS.491.4591E, 2021MNRAS.503.4075G, 2021MNRAS.505...18E, 2022MNRAS.517.1398B}, but the inner density cusp of a subhalo remains largely intact, which is crucial for DM annihilation studies because the luminosity is dominated by the dense central regions. This effect typically reduces the boost by $20$–$30\%$ and is modulated by the environment, with filament haloes generally dominating at high masses, walls intermediate, and voids the most suppressed.

With the distance- and environment-dependent subhalo concentrations and tidal stripping included, the resulting boosts in Fig.~\ref{fig:boost_moline_tidal} follow the same qualitative trends as Fig.~\ref{fig:Boost_scp}. Filament haloes are slightly more enhanced at high host \mvir{} masses, reaching $\sim 1.12$ times the cosmic mean compared to $\sim 1.07$ in the \citetalias{2014MNRAS.442.2271S}-based environment-dependent model, while walls and voids retain similar intermediate and suppressed behavior, respectively. This illustrates that incorporating distance-dependent concentrations amplifies the boost in dense cosmic web environments without altering the overall pattern set by the \citetalias{2014MNRAS.442.2271S}-based environment-dependent model.

For convenient implementation in future semi-analytical models or N-body analyses, we provide an analytical representation of the environment-specific boost factors shown in Fig.~\ref{fig:boost_moline_tidal}. Following the general functional form of \citetalias{2017MNRAS.466.4974M}, we parametrize these values using a logarithmic polynomial expansion:
\begin{equation}
\log_{10} B(M_{200}, z=0) = \sum_{i=0}^{3} b_i 
\left[ \log_{10}\left(\frac{M_{200}}{M_\odot}\right) \right]^i ,
\label{eq:moline_poly}
\end{equation}
where the environment-dependent coefficients $b_i$ are listed in Table~\ref{tab:boost_poly_coeffs}.
The uncertainties are derived from the fit covariance matrix. This environment-dependent extension of the \citetalias{2017MNRAS.466.4974M} framework is intended as a practical fitting prescription for downstream applications, including annihilation forecast calculations, halo-population modeling, and environment-aware semi-analytic studies of dark-matter targets.

The fit reported above is calibrated for $z=0$ host haloes in the mass range $10^9\simlt M_{200}/(M_\odot/h)\simlt10^{13}$, and for the three large-scale environments considered in this work, namely filaments, walls, and voids. It should therefore be used as an interpolation formula within this calibrated domain. Its validity outside this mass range, at higher redshift, or for alternative cosmological calibrations remains to be tested. Across the calibrated mass range, the polynomial representation reproduces the boost curves very closely, with absolute residuals typically of order $10^{-2}$ and nowhere exceeding about $2\times 10^{-2}$.
\section{Summary and conclusions \label{sec:summary}}
In this work, we have explored how the cosmic web environment influences DM haloes and their substructure, and the potential implications for annihilation signals. Using the high-resolution \textsc{COLOR} simulation \citep{Hellwing2016}, together with the \citetalias{2014MNRAS.442.2271S} semi-analytic framework and the environment-dependent trends calibrated in \citet{Hellwing2021,2025A&A...700A..65M}, we examined how host-halo concentration, subhalo abundance, and subhalo internal structure vary systematically across filaments, walls, and voids. These environment-dependent corrections were incorporated into the computation of subhalo boost factors, and we also considered the distance- and environment-dependent concentration model of \citetalias{2017MNRAS.466.4974M} to explore the impact of tidal stripping and subhalo radial positions on the boosts.

The main findings of this study are:
\begin{itemize}
    \item Host halo concentrations show a dependence on the cosmic web environment, with haloes in filaments generally more concentrated, walls intermediate, and void haloes less concentrated relative to the cosmic mean.
    
    \item The subhalo mass function varies systematically with environment. Filament haloes host the largest number of subhalos at higher host masses, whereas wall haloes can have slightly higher subhalo numbers at lower host masses. Void haloes consistently contain the fewest subhalos.

    \item Subhalo internal densities, quantified via $\rmax$ ratios or characteristic concentrations, also vary with environment. Subhalos in filaments tend to have higher densities, those in walls are intermediate, and void subhalos are lower in density, reflecting differences in formation times and tidal influences.  

    \item Environment-modulated subhalo boost factors reflect the combined effects of host-halo concentration, subhalo abundance, and subhalo internal structure. Filament haloes consistently exhibit the largest boosts, wall haloes remain intermediate, and void haloes are the most suppressed. Relative to the cosmic mean, filament boosts rise from about $0.85$ at low host masses to $1.07$ near $10^{11}\,\Msun$, wall boosts remain below unity and decline to $\sim0.76$ at the high-mass end, while void boosts stay suppressed across the full mass range and can drop to approximately $0.70$.

    \item Incorporating the distance-dependent concentration model of \citetalias{2017MNRAS.466.4974M} preserves the same environmental ordering, but increases the high-mass filament boost from $\sim1.07$ in the baseline \citetalias{2014MNRAS.442.2271S}-based model to $\sim1.12$. At low host masses, filament boosts remain below unity, while wall and void boosts stay intermediate and suppressed, respectively.
\end{itemize}

These results suggest that the predicted DM annihilation signal is influenced not only by the number of subhalos but also by their internal densities, which vary with environment. The observed modulation implies that different cosmic web locations can lead to subtle but meaningful variations in expected signals. Including environmental effects alongside structural variations provides a framework that allows for more nuanced predictions of annihilation boosts.

Although this study is limited to two hierarchical levels of subhalos, it illustrates how variations in halo and subhalo properties may influence DM annihilation signals. The principal environmental modulation of the boost remains present across the different subhalo and concentration prescriptions explored, although the detailed amplitude and low-mass environmental ordering can depend on the treatment of nested substructure. Future work could consider additional complexities, such as tidal interactions, halo assembly histories, and other local environmental factors, to further assess their potential impact on predicted signals.

A separate limitation concerns the extrapolation of halo-profile self-similarity down to \(M_{\min}\simeq10^{-6}\,M_\odot\). Recent simulations suggest that microhaloes near the free-streaming scale may form through early collapse of density peaks and develop compact inner structures, often described as prompt cusps with \(\rho\propto r^{-1.5}\) \citep{2020MNRAS.492.3662I,2023JCAP...10..008D}. Such cusps may survive hierarchical evolution in some cases and enhance the absolute annihilation boost \citep{2026arXiv260119863A}. This regime lies beyond the resolved calibration domain of the present work. Since the cosmic web dependence of prompt-cusp formation and survival has not yet been established, our conclusions should be interpreted primarily as predictions for the relative environmental modulation of the boost, rather than as a recalibration of its absolute normalization at the free-streaming scale.

Overall, incorporating variations in host halo concentration, subhalo abundance, and internal density structure associated with the cosmic web provides a more detailed understanding of how environmental factors might shape the DM annihilation signal. These findings demonstrate the potential importance of considering environmental context when interpreting indirect detection signals or constructing theoretical predictions.
\subsection{Implications for Indirect Detection Signals}

In our model, the cosmic web environment modulates boost factors by tens of percent: the ratio $B(M, \text{env})/B_{\rm CM}(M)$ ranges from approximately $70\%$ in voids to $112\%$ in filaments across the host-mass range explored (see Fig.~\ref{fig:boost_moline_tidal}). This level of variation is comparable to other key theoretical uncertainties, such as halo-to-halo scatter in concentration and uncertainties in the normalization and slope of the SHMF, demonstrating that environmental effects are a significant source of variation rather than a minor higher-order correction.

The sensitivity arises from the coordinated variation of host halo concentration, subhalo abundance, and internal density structure. Because annihilation rates scale as $\rho^2$, even modest shifts in these properties across the cosmic web can produce discernible differences in the expected DM signal. Consequently, haloes residing in denser, more massive regions, such as filaments, are predicted to generate stronger annihilation signatures than their counterparts in underdense environments. This is consistent with foundational substructure studies \citep[e.g.,][]{2008ApJ...686..262K, 2008Natur.454..735D, 2019Galax...7...68A}, while adding an explicit spatial context: the increased subhalo mass fraction and elevated central densities in filaments imply that annihilation luminosities can be systematically enhanced relative to an environment-agnostic, cosmic-mean population.

For observational strategies, these results suggest that targets located in different large-scale environments may exhibit modest variations in expected annihilation flux depending on their position within the cosmic web.
Selecting haloes in filamentary regions could yield slightly higher expected annihilation signals, whereas searches in under-dense voids may be subject to lower expected fluxes. Similarly, for studies of the Milky Way or nearby galaxies, acknowledging that the local environment shapes the subhalo population could lead to more nuanced foreground or background estimates. Including information on the spatial distribution and structural variation of subhalos allows for a refinement of predictions without requiring drastic changes to existing detection paradigms.

Comparisons with alternative subhalo concentration models, such as distance-dependent prescriptions, indicate that variations in subhalo internal density can further alter the boost factor, particularly for specific host halo mass ranges. While the trends observed here are consistent with general substructure expectations, they suggest that environmental modulation introduces an additional layer of variation that is highly relevant for interpreting indirect detection signals. In practice, accounting for the cosmic web environment provides a more nuanced view of potential annihilation luminosities, helping to quantify relative differences between haloes rather than producing absolute predictions. This approach assists in refining theoretical models and observational strategies while maintaining consistency with existing expectations.

Beyond the final boost-factor predictions, this work also provides a set of intermediate, environment-dependent ingredients that may be useful more broadly: corrections to the host-halo concentration, the differential SHMF, and the subhalo internal-structure proxy based on the $V_{\max}$--$R_{\max}$ relation. Because these components are fitted separately and expressed in modular form, they can in principle be propagated into other semi-analytic or forward-model frameworks that are sensitive to subhalo abundance and internal structure. This includes strong-lensing mass modelling, where forward modelling of lensed images is increasingly used to constrain the inner mass distribution of lenses and the presence of low-mass dark subhaloes \citep{2025MNRAS.544..782H}. While such an application is beyond the scope of the present paper, the fitted ingredients derived here provide a natural starting point for incorporating cosmic web environmental information into lensing-oriented subhalo models.

 \begin{acknowledgments}
    The authors thank the anonymous referee for valuable comments and suggestions that have helped to improve the paper. The authors thank Oliver Newton for discussion and comments at the early stage of this project and Krishna Naidoo for running the \cactus{} cosmic web classification for us on the \textsc{COLOR} data. This work was supported via the research project "COLAB" funded by the National Science Centre, Poland, under agreement number UMO-2020/39/B/ST9/03494. 
 \end{acknowledgments}
\appendix
\section{Subhalo Concentration Model}
\label{app:moline_concentration}
The subhalo concentration model described here is introduced in Section~\ref{sec:results}, where its application to computing the subhalo boost factor is presented. This section provides the full parametrization of the model, including the dependence on subhalo mass and radial position within the host halo.

We adopt the distance-dependent concentration model of \citetalias{2017MNRAS.466.4974M}, calibrated against high-resolution numerical simulations. In this framework, the concentration depends explicitly on both the subhalo mass and its radial distance from the host center. This formulation incorporates the combined effects of hierarchical structure formation and tidal evolution inside the host halo.

The concentration is parametrized as Eq.~(7) in \citetalias{2017MNRAS.466.4974M}:
\begin{equation}
\begin{aligned}
c_{\mathrm{200}}(m_{\mathrm{200}}, x_{\rm sub}) &= c_0 
\Bigg[ 1 + \sum_{i=1}^{3} a_i 
\left( \log_{10} \frac{m_{\mathrm{200}}}{10^8 \Msun} \right)^i \Bigg] \\
&\quad \times \Big[ 1 + b \, \log_{10} (x_{\rm sub}) \Big],
\end{aligned}
\end{equation}
where $c_0 = 19.9$, $a_i = \{-0.195, 0.089, 0.089\}$, and $b = -0.54$. The radial coordinate is defined as $x_{\rm sub} = R_{\rm sub}/R_{\rm vir}$. The first bracket captures the mass dependence, reflecting the earlier formation times of lower-mass haloes in a hierarchical scenario. The second bracket introduces a radial modulation that decreases $c_{\rm 200}$ with increasing distance from the host center, accounting for tidal stripping and dynamical processing. As a result, subhalos located closer to the host center are predicted to be more concentrated than those in the outskirts.

\begin{figure*}
  \centering
  \includegraphics[width=\linewidth]{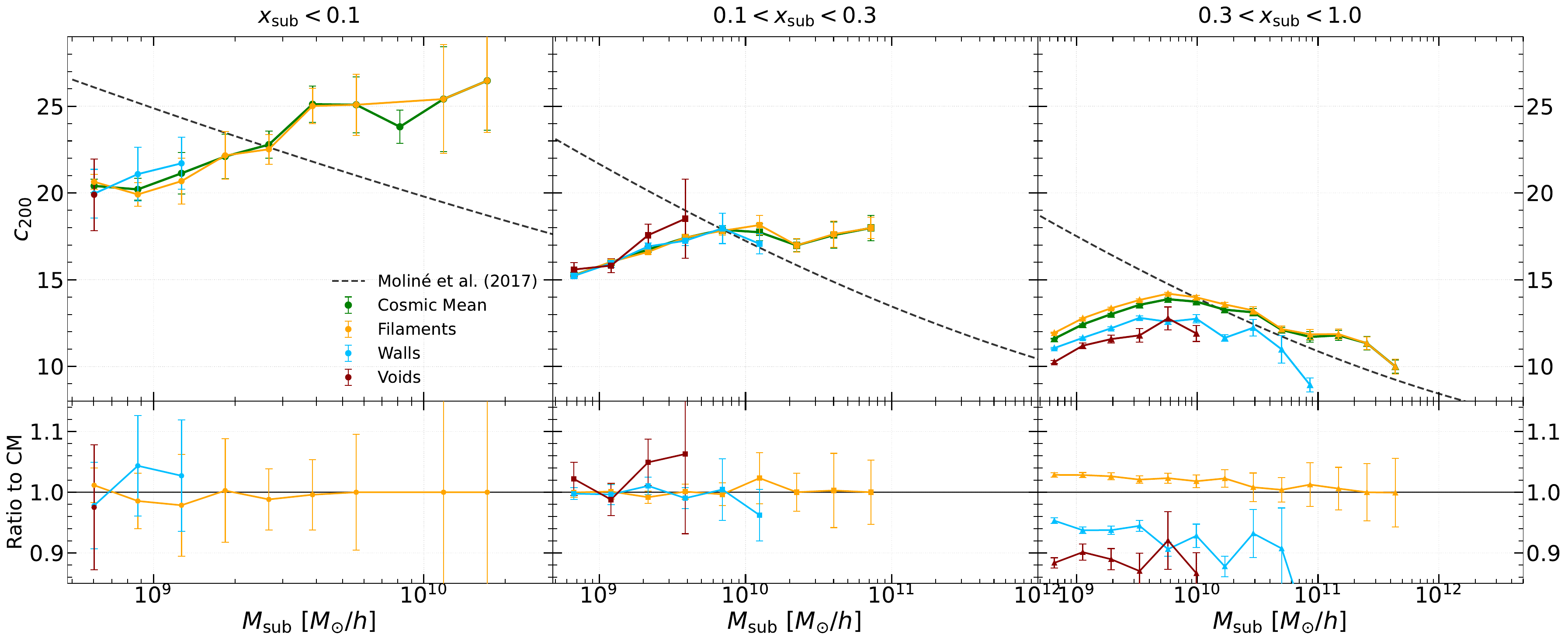}
  \caption{Median subhalo concentration $c_{\rm 200}$ as a function of subhalo mass $m_{\rm sub}$ in three radial bins. In each panel, the upper sub-panel displays the median concentration for different large-scale environments, with error bars indicating the $1\sigma$ bootstrap uncertainties, while the black dashed line represents the concentration-mass relation from \citetalias{2017MNRAS.466.4974M} evaluated at the median $x_{\rm sub}$ of each bin. The lower subpanel shows the concentration in each environment relative to the cosmic mean.}
\label{fig:c200_moline}
\end{figure*}
The environmental dependence of subhalo concentrations is illustrated in Fig.~\ref{fig:c200_moline}. Each panel corresponds to a different radial bin, with the upper subpanel showing the median $\cvir$ as a function of $m_{\rm 200}$ in distinct large-scale environments. The lower sub-panel shows the concentration in each environment relative to the cosmic mean, indicating the effect of the cosmic web on subhalo internal structure.

The influence of the cosmic web on subhalo concentration is secondary to radial position, with a negligible impact for subhalos in the inner host regions ($x_{\rm sub} < 0.1$). However, a distinct environmental hierarchy emerges in the outer radial bins ($0.3 < x_{\rm sub} < 1.0$). In these outskirts, subhalos in filaments closely track the cosmic mean, effectively defining the average concentration. In contrast, subhalos in voids exhibit the most significant deviation, with concentrations approximately 12--15\% lower than the cosmic mean. Subhalos residing in walls represent an intermediate population, showing a modest deficit of roughly 5\% compared to the mean. This suggests that less dense cosmic environments correlate with lower subhalo concentrations as one moves toward the host outskirts.

This behavior is consistent with tidal evolution within the host potential. Subhalos in the inner regions experience stronger tidal stripping and repeated dynamical interactions, which effectively homogenize their structural properties and erase differences inherited from the large-scale environment. In contrast, subhalos in the outskirts are less dynamically processed and therefore retain a residual dependence on the cosmic web, with the observed trends reflecting the local density conditions and assembly histories of their respective environments before accretion.

\bibliography{bibliography}
\bibliographystyle{apsrev4-2}
\end{document}